# Optimal Rate Irregular LDPC Codes in Binary Erasure Channel

H. Tavakoli[1], M. Ahmadian[2], M. Reza Peyghami[3]

**Abstract**

In this paper, we design the optimal rate capacity approaching irregular Low-Density Parity-Check code ensemble over Binary Erasure Channel, by using practical Semi-Definite Programming approach. Our method does not use any relaxation or any approximate solution unlike previous works. Our simulation results include two parts; first, we present some codes and their degree distribution functions that their rates are close to the capacity. Second, the maximum achievable rate behavior of codes in our method is illustrated through some figures.

***Index***: *LDPC code, Infinite analysis method, Density evolution, Semi-definite programming, Linear programming*

**I. Introduction**

Gallager introduced Low Density Parity Check (LDPC) codes in the early sixties [1,2]. After three decades of Gallager's work, MacKay and Neal's rediscovery [3] attracted more attention to LDPC codes. Approaching the channel capacity is one of the important properties of long irregular LDPC codes [4].

One of the classical channel models is Binary Erasure Channel (BEC), presented by Elias in 1955 [5]. Nowadays, BEC has become popular as a communication model in packet-loss networks [6]. In this paper, the simplest model for the capacity-approaching problem that is related to BEC is considered. This simplicity comes from Density Evolution (DE) constraint which was introduced by Richardson and Urbanke in [4] and [7-8]. DE is a numerical method for understanding how the iterative message passing decoder works in infinite number of iterations. Both infinite and finite length considerations for LDPC codes over BEC have been studied in [9-13].

[1] Electrical Engineering Department, K.N. Toosi University of Technology, Tehran, Iran. Email: tavakoli@ee.kntu.ac.ir
[2] Electrical Engineering Department, K.N. Toosi University of Technology, Tehran, Iran. Email: m_ahmadian@kntu.ac.ir
[3] Mathematics Department, K.N. Toosi University of Technology, Tehran, Iran. Email: peyghami@kntu.ac.ir

However, finding an efficient method for approaching the channel capacity in BEC has been remained as an open problem. Approaching the channel capacity is based on the structure of the code, although some experiences of achieving the channel capacity in infinite node degree distribution have been reported in [14-16]. Degree distributions of a code illustrate the behavior of approaching the channel capacity. Some attempts for finding proper degree distributions are presented in references [9,17,18,19] but the methods have not been efficient to approach the channel capacity.

In [20], instead of solving the optimization problem with its non-linear constraint, the main problem of approaching capacity is presented. In addition, some other constraints on the degree distributions over Memory-less Binary-Input Output-Symmetric (MBIOS) are introduced. However, the problem of finding good degree distribution was not solved in that paper.

There are four ways for finding good degree distributions. In the first method, the main algorithm is based on an evolutionary optimization method, such as hill climbing, genetic algorithm [4, Section IV.]. It is known that these heuristic methods suffer from some disadvantages, such as: 1) not guarantee a feasible answer 2) not converging 3) being sensitive to their subroutine and start point. These disadvantages restrict their use in other classes of problems.

In the second method, finding a direct way towards the answer such as differential evolution is the main target of these algorithms [4, Section IV.]. These types of algorithms are based on using infinite number of iterations of an optimization method which may have a loop without considering convergence or certification for the optimal answer.

In the third approach, for finding good degree distribution, which approaches the channel capacity, using Linear Programming (LP) method is the main idea [9, Section V.]. It is known that for maximizing the rate, the optimization problem is:

$$\text{Min} \sum \frac{\rho_j}{j}$$

Subject to: $\sum \rho_j(1 - \varepsilon\lambda(x_i)) > 1 - x_i$

where $\rho(x)$ and $\lambda(x)$ are check and variable nodes degree distributions, respectively and $x_i$s, $1 < i < N$, are a set of some fractional values in (0,1]. So a LP problem for finding $\rho_j$s provided that $\lambda_i$s can be defined.

In the fourth approach which is developed in this paper, using Semi-Definite Programming (SDP) is the main idea. In this method, instead of using some samples of non-linear DE constraint like the pervious methods, the whole space is considered [21,22]. Whereas relaxation of some constraints in the optimization problem leads to a sub-optimal solution, in our method which is based on an exact constraint with no relaxation, the solution of SDP problem would be certainly optimal.

Some other methods for designing good degree distributions without using optimization method are presented in [17] and [18]. The basic idea of these papers is based on Taylor's series of $\rho^{-1}(1-x)$ in infinite way.

According to [23], finding maximum code rate under a fixed ensemble is an important task where some ordered bounds for fixed parity check degree distribution and erasure probability are represented. Our method achieves these bounds, as well.

The organization of the paper is as follows. In Section II, we provide a brief background on defining the main problem for optimizing degree distribution. In Section III, we describe SDP reformulation for optimal rate problem. In Section IV, we introduce how we can optimize a code. Finally, in Section V, we illustrate our contribution with simulation results. Section VI summarizes and concludes the paper.

## II. Problem Definition

In this paper, for achieving maximum rate for a code, we focus on irregular LDPC code ensemble over BEC. Let $H$ be the parity check matrix of a code with a bipartite graph with maximum variable node degree $D_v$ and check node degree $D_c$. A code is represented with edge degrees specified by two polynomials:

$$\rho(x) = \sum_{j=2}^{D_c} \rho_j x^{j-1} \qquad \lambda(x) = \sum_{i=2}^{D_v} \lambda_i x^{i-1} \tag{1}$$

The coefficients of each polynomial denote the probability of having a variable or check node with its index degree, i.e.,

$$\sum_{j=2}^{D_c} \rho_j = 1 \qquad \sum_{i=2}^{D_v} \lambda_i = 1 \qquad \lambda_i \geq 0, \rho_j \geq 0 \tag{2}$$

It is well known that the design rate of a code is defined as [8]:

$$R = 1 - \frac{\sum_{j=2}^{D_c} \rho_j/j}{\sum_{i=2}^{D_v} \lambda_i/i} \tag{3}$$

For a BEC with given erasure probability $\varepsilon > 0$, the capacity is $C = 1 - \varepsilon$. In BEC with given degree distribution, the necessary and sufficient conditions for achieving the zero error probability in infinite number of iterations for message passing decoder which comes from DE is [8]:

$$\lambda(1 - \rho(1-x)) \leq \frac{x}{\varepsilon} \qquad \forall x \epsilon [0, \varepsilon] \tag{4}$$

It is clear that

$$R = 1 - \left(\sum_{j=2}^{D_c} \rho_j/j\right) / \left(\sum_{i=2}^{D_v} \lambda_i/i\right) \leq C = 1 - \varepsilon,$$

Thus, $\varepsilon \leq \left(\sum_{j=2}^{D_c} \rho_j/j\right) / \left(\sum_{i=2}^{D_v} \lambda_i/i\right)$. For approaching the channel capacity, the inequality must tend to the equality. In order to maximize the rate of the channel while achieving zero error probability, it is sufficient to solve any of the following Non-Linear Programming (NLP) optimization problems:

- NLP1) The check node degree and erasure probability are fixed and variable node degree is found accordingly, i.e.:

$$\text{NLP1: Max } \sum \frac{\lambda_i}{i}$$

$$\text{Subject to: } \lambda_i \geq 0$$

$$\sum \lambda_i = 1$$

$$\sum \lambda_i \bigl(1 - \rho(1 - \varepsilon x)\bigr)^{i-1} \leq x \quad \forall x \in (0,1]$$

- NLP2) The variable node degree and erasure probability are fixed and check node degree is found accordingly, i.e.:

$$\text{NLP2: Min } \sum \frac{\rho_i}{i}$$

$$\text{Subject to: } \rho_i \geq 0$$

$$\sum \rho_i = 1$$

$$\sum \rho_i \bigl(1 - \varepsilon \lambda(1 - x)\bigr)^{i-1} \leq x \quad \forall x \in (0,1]$$

- NLP3) The variable node degree and the check node degree are fixed and erasure probability must be found for holding the equality.

$$\text{NLP3: Min } t$$

$$\text{Subject to: } \lambda_i \geq 0, \ t \geq 1$$

$$\sum \lambda_i = 1$$

$$t = \frac{1}{\varepsilon}$$

$$\sum \lambda_i \bigl(1 - \rho(1 - x)\bigr)^{i-1} \leq tx \quad \forall x \in (0,1]$$

These problems are semi-infinite optimization problems, i.e., they include infinite number of constraints. One way for solving these problems is to discretize the problem by partitioning the continuous interval $(0,1]$ for $x$ to discrete set $\{x_0, x_1, ..., x_N\} \subseteq (0,1]$, [9, Section V.]. In this approach only some points in continuous interval are considered and other points are ignored. Therefore, the problem is converted to an optimization problem with finite number of constraints. However, the cost that has been paid for this discretization is that a sub-optimal solution is achieved.

Considering the NLP1, NLP2 and NLP3, we realize that the main part that makes them to be semi-infinite problems is their last constraint. Due to this fact, we reformulate these constraints as a Linear Matrix Inequality (LMI) and, therefore, we get a semi-definite reformulation for NLP1. Other non-linear programming problems NLP2 and NLP3, can be reformulated in a similar method as NLP1. This reformulation helps us to solve the problem by using polynomial time interior-point methods. It is notable that this reformulation leads us to get an exact solution instead of suboptimal solutions. Numerical results confirm our claim in comparison with the existing methods.

## III. SDP reformulation for the optimal rate problem

In this section, we first introduce some notations and definitions, and then by using some lemmas, we show the trajectory of the optimization way which will be completed in the next section.

### A. Definitions

The definitions and notations provided in this sub-section are used throughout the paper. All of these concepts are the standard definitions and notations in the context of conic optimization and have been selected from [24].

**Definition 1:** Let $S^k$ be the set of all symmetric $k \times k$ matrices and $S_+^k$ be a sub-set of $S^k$ with positive semi-definite matrices. We use the notation $A \succcurlyeq 0$ when the matrix $A$ is positive semi-definite, i.e, for all $x \in \mathbb{R}^k$, we have $x^T A x \geq 0$.

**Definition 2:** The Lorentz cone (Ice-Cream Cone) $\mathcal{L}^k$ is defined as:

$$\mathcal{L}^k = \left\{ x \in \mathbb{R}^k \middle| x_k \geq \sqrt{x_1^2 + x_2^2 + \cdots + x_{k-1}^2} \right\} \tag{5}$$

We simply write $x \succcurlyeq_{\mathcal{L}^k} 0$ for $x \in \mathcal{L}^k$. In general, we write $X \succcurlyeq_{\mathcal{K}} 0$ when $X$ belongs to a certain cone $\mathcal{K}$.

**Definition 3:** A set $X \subseteq \mathbb{R}^n$ is called Semi-Definite representable (SDr), if there exists an affine map $\mathcal{A}(x, u)$ from $\mathbb{R}^n \times \mathbb{R}^m$ to $S^k$, with extra variables $u \in \mathbb{R}^m$ such that:

$$X = \{x | \exists u : \mathcal{A}(x, u) \succcurlyeq 0\} \tag{6}$$

In fact,

$$\mathcal{A}(x, u) = \sum_j x_j A_j + \sum_l u_l B_l + C \tag{7}$$

where $A_j$s, $B_l$s and $C$ are symmetric matrices. We call such an explicit representation of $X$ as Semi-Definite Representation (SDR) for $X$.

### B. Problem Formulation

Now, we describe our reformulation in details. To do so, we focus on the main non-linear constraints of the NLP1, i.e,

$$\lambda(1 - \rho(1 - \varepsilon x)) \leq x \quad \forall x \in [0,1] \tag{8}$$

In other words, the feasible region of NLP1 contains all vectors $\lambda$ that satisfy the following equation:

$$P(x) = x - \lambda(1 - \rho(1 - \varepsilon x)) \geq 0 \quad \forall x \in [0,1] \tag{9}$$

Since $\lambda(x)$ and $\rho(x)$ are polynomials, the function $P(x)$ is a polynomial function with degree at most $D_c D_v$. Let $q = D_c D_v$ and

$$P(x) = \sum_{j=1}^{q} p_j x^j \tag{10}$$

where $p_j = p_j(\lambda_1, \lambda_2, \ldots, \lambda_{D_v}, \rho_1, \rho_2, \ldots, \rho_{D_c}, \varepsilon)$. For $\rho(x) = x^n$, we have the following technical lemmas for computing the coefficients $p_j$, $1 \le j \le D_c$.

**Lemma 1**: Let $F(x) = (a_1 x + a_2 x^2 + \cdots + a_n x^n)^{k-1}$ be written as:

$$F(x) = \sum_{m=k-1}^{n(k-1)} b_m x^m \tag{11}$$

Then, we have:

$$b_m = \sum_{i_l \in \mathbb{N} \cup \{0\}, \sum_{l=1}^{k-1}(l\, i_l)=m, \sum_{l=1}^{n} i_l = k-1} \binom{k-1}{i_1, i_2, \ldots, i_n} \prod_{l=1}^{n} a_l^{i_l}. \tag{12}$$

**Proof**: See [25].

**Lemma 2**: Let $\rho(x) = x^n$ and $P(x)$ be defined as in (8) and (9). Then, we have:

$$p_j = \begin{cases} 1 - \lambda_2 n\varepsilon & j = 1 \\ -\sum_{l=2}^{j+1} \lambda_l \varphi_{j,l-1} & j \ge 2 \end{cases} \tag{13}$$

where the coefficients $\varphi_{k,i-1}$ are defined as:

$$\varphi_{k,i-1} = (-1)^{k+i} \varepsilon^k \sum_{\pi_l \ge 0: \sum_{l=1}^{i-1} \pi_l = k} \binom{n}{\pi_1}\binom{n}{\pi_2} \cdots \binom{n}{\pi_{i-1}} \tag{14}$$

**Proof**: Considering $\rho(x) = x^n$, we have:

$$P(x) = \sum_{j=1}^{q} p_j x^j = x - \lambda(1 - (1 - \varepsilon x)^n) = x - \sum_{i=2}^{D_v} \lambda_i (1 - (1 - \varepsilon x)^n)^{i-1}$$

$$= x - \sum_{i=2}^{D_v} \lambda_i \left(n\varepsilon x - \binom{n}{2}\varepsilon^2 x^2 + \binom{n}{3}\varepsilon^3 x^3 - \cdots + (-1)^{n+1}\varepsilon^n x^n\right)^{i-1} = x - \sum_{i=2}^{D_v} \lambda_i \Psi_{i-1}$$

where

$$\Psi_{i-1} = \left(n\varepsilon x - \binom{n}{2}\varepsilon^2 x^2 + \binom{n}{3}\varepsilon^3 x^3 - \cdots + (-1)^{n+1}\varepsilon^n x^n\right)^{i-1}.$$

Assume that $\Psi_{i-1} = \sum_{k=i-1}^{n(i-1)} \varphi_{k,i-1} x^k$. Then, using Lemma 1, we have

$$\varphi_{k,i-1} = \left(-\sum_{i=1}^{n}\binom{n}{i}(-\varepsilon x)^i\right)^{k-1} = \left(n\varepsilon x - \binom{n}{2}\varepsilon^2 x^2 + \binom{n}{3}\varepsilon^3 x^3 \mp \cdots + (-1)^{n+1}\varepsilon^n x^n\right)^{k-1}$$

which completes the proof of lemma.

In order to express (9) as LMI, we first consider its general form, i.e.,

$$P(x) \geq 0, \quad \forall x \in \mathbb{R}. \tag{15}$$

We first show that this constraint can be stated as LMI. The following lemma proves that the affine image preserves Semi-Definite representability (SDr) [24].

**Lemma 3**: Let $X \subseteq \mathbb{R}^n$ be a SDr set, and $F: X \to \mathbb{R}^k$ be an affine map, i.e.,

$$F(x) = Ax + b \tag{16}$$

Then, the set

$$Image(X) = \{y | \exists x \in X;\ y = Ax + b\} \tag{17}$$

is also SDr set.

**Proof**: Since $X$ is a SDr set, we have $X = \{x|\ \exists u: \mathcal{A}(x, u) \succcurlyeq 0\}$, where $\mathcal{A}(x, u)$ is an affine map. Therefore, we obtain

$$Image(X) = \{y | \exists (x, u);\ \mathcal{A}(x, u) \succcurlyeq 0, y = Ax + b\}$$

$$= \{y | \exists (x, u);\ \mathcal{B}(y, (x, u)) \succcurlyeq_\mathcal{K} 0\} \tag{18}$$

where

$$\mathcal{B}(y,(x,u)) \equiv \begin{bmatrix} \mathcal{A}(x,u) \\ \hline 0 \\ (y - Ax - b)_1 \\ \hline 0 \\ (Ax + b - y)_1 \\ \hline \vdots \\ \hline 0 \\ (y - Ax - b)_k \\ \hline 0 \\ (Ax + b - y)_k \end{bmatrix}$$

and $\mathcal{K} = S \times L^2 \times ... \times L^2$ is a direct product of semi-definite cone $S$ with *2k* times Lorentz cone $\mathcal{L}^2$.

It can be easily proved that the inverse *Image* of a SDr set under a linear transformation is also SDr [24]. Let $P_{2k}^+(\mathbb{R})$ be the set of coefficients of all polynomials with $degree \leq 2k$ which are non-negative on the entire axis. The following lemma shows that the set $P_{2k}^+(\mathbb{R})$ is SDr and provides a Semi-Definite Representation (SDR) for this set [24].

**Lemma 4**: The set $P_{2k}^+(\mathbb{R})$ is SDr.

**Proof**: Consider the following map from $S_+^{k+1}$ on $P_{2k}^+(\mathbb{R})$:

$$F: S_+^{k+1} \to P_{2k}^+(\mathbb{R}) \tag{19}$$

$$F(B) = b = (b_0, b_1, ... b_{2k})$$

where $b_l = \sum_{i+j=l} B_{ij}$. Now, we prove that $F$ is an affine map. To do so, it is sufficient to show that the following equality holds for all $B, C \in S_+^{k+1}$ and $\alpha \in \mathbb{R}$:

$$F(\alpha B + C) = \alpha F(B) + F(C). \tag{20}$$

We have:

$$F(\alpha B + C) = \sum_{i+j=l}(\alpha B_{ij} + C_{ij}) = \sum_{i+j=l}\alpha B_{ij} + \sum_{i+j=l}C_{ij} = \alpha\sum_{i+j=l}B_{ij} + \sum_{i+j=l}C_{ij}$$

$$\Rightarrow F(\alpha B + C) = \alpha F(B) + F(C).$$

Therefore, due to Lemma 3, the set $P_{2k}^+(\mathbb{R})$ is SDr. Note that the linear map $F$ provides the following SDR for $P_{2k}^+(\mathbb{R})$:

$$P_{2k}^+(\mathbb{R}) = \left\{ b = \{b_0, b_1, \ldots b_{2k}\} \;\middle|\; \exists\, B \in S_+^{k+1}; \; b_l = \sum_{i+j=l} B_{ij} \right\}$$

Lemma 5 and 6 helps us to convert the constraints of (9) to linear matrix inequalities.

**Lemma 5**: $P_q^+([0,\infty))$, the set of coefficient of all polynomials with *degree* $\leq q$ that are non-negative on the non-negative ray, is SDr.

**Proof**: Indeed, $P_q^+([0,\infty))$ is the inverse image of the SDr set $P_{2q}^+(\mathbb{R})$ under the linear transformation of coefficients of polynomials induced by the mapping:

$$P(x) \mapsto \Pi(x) = P(x^2) : \mathcal{P}_q \mapsto \mathcal{P}_{2q} = \Pi \qquad (21)$$

where $\mathcal{P}_l$ is the space of algebraic polynomials of *degree* $\leq l$. Now, using Lemma 3, the proof would be straightforward.

**Lemma 6**: The set $P_q^+([0,1])$, of coefficients of all polynomials with *degree* $\leq q$ that are non-negative on the interval $[0,1]$, is SDr.

**Proof**: Indeed, $P_q^+([0,1])$ is the inverse image of the SDr set $P_{2q}^+(\mathbb{R})$ under the linear transformation of coefficients of polynomials induced by the mapping:

$$P(x) \mapsto \Pi(x) = (1+x^2)^q P\left(\frac{x^2}{1+x^2}\right) : \mathcal{P}_q \mapsto \mathcal{P}_{2q} = \Pi. \qquad (22)$$

Using the above lemmas, we provide an LMI formulation for the constraint of (9) in the next section.

## IV. Code Optimization

In this section, we provide an explicit semi-definite representation for the set $\{\lambda | P(x) \geq 0, \forall x \in [0,1]\}$. In fact, we would like to replace the infinite constraints of (9) by the intersection of affine constraints and some finite LMIs. This leads us to solve NLP1 taking full responsibility for all of the constraints instead of ignoring some of these constraints by discretizing the interval $(0,1]$ to finite points, as it has been considered in the Literature e.g. [9]. The following Lemma and Theorem lead us to the aim of this section.

**Lemma 7**: Let $\Pi(x) = (1+x^2)^q P\left(\frac{x^2}{1+x^2}\right) = \sum_{j=0}^{2q} \Pi_j x^j$, where $P(x)$ is defined as (9). Then, we have:

$$\Pi_t = \begin{cases} \sum_{i=1}^{j} \binom{q-i+1}{j-i+1} p_{i-1} & t = 2j \\ 0 & t = 2j+1 \end{cases} \quad (23)$$

**Proof:** We have:

$$\Pi(x) = \sum_{j=0}^{q} p_j x^{2j}(x^2+1)^{q-j} \quad (24)$$

Using Newton's expansion, we have:

$$x^{2j}(x^2+1)^{q-j} = x^{2j} \sum_{r=0}^{q-j} \binom{q-j}{r} x^{2r} = \sum_{r=0}^{q-j} \binom{q-j}{r} x^{2r+2j} =$$

$$\binom{q-j}{0} x^{2j} + \binom{q-j}{1} x^{2+2j} + \binom{q-j}{2} x^{4+2j} + \cdots + x^{2q} \quad (25)$$

Therefore, we obtain:

$$\Pi(x) = \sum_{j=1}^{q} \left\{ p_j \binom{q-j}{0} x^{2j} + p_j \binom{q-j}{1} x^{2+2j} + p_j \binom{q-j}{2} x^{4+2j} + \cdots + p_j x^{2q} \right\} \quad (26)$$

And we have:

$$\Pi_t = \begin{cases} \sum_{i=1}^{j} \binom{q-i+1}{j-i+1} p_{i-1} & t = 2j \\ 0 & t = 2j+1 \end{cases}$$

**Theorem 1**: Let $\Pi(x)$ be defined as in Lemma 7. Then, NLP1 is equivalent to the following semi-definite programming problem:

---

$SDP1:$  $Max \; \sum \frac{\lambda_i}{i}$

$\quad\quad Subject\; to: \quad \sum \lambda_i = 1$

$\quad\quad\quad\quad\quad\quad\quad\quad \Pi_l = \sum_{i+j=l} B_{ij}, \quad\quad 0 \le l \le 2q$

$\quad\quad\quad\quad\quad\quad\quad\quad B \succcurlyeq 0, \quad 0 \le \lambda_i \le 1$

---

**Proof**: According to the discussions of pervious section, the vector $\lambda$ satisfies (4) if and only if its image by affine mapping $P(x) \to \Pi(x) = (1+x^2)^q P\left(\frac{x^2}{1+x^2}\right)$ from $\mathbb{R}$ to $(0,1]$ satisfies $\Pi(x) \ge 0$, for all $x \in \mathbb{R}$. Using (19), this equality holds if and only if there exists a symmetric positive semi-definite matrix $B = (B_{ij})_{(q+1)\times(q+1)}$ so that it satisfies the following equations:

$$\begin{cases} \Pi_l = \sum_{i+j=l} B_{ij}, & 0 \le l \le 2q \\ B \succcurlyeq 0, \end{cases} \tag{27}$$

The proof is completed by replacing these systems of linear equations and LMIs in NLP1. □

In order to illustrate these results, we provide a simple structured example to show how these results can be handled in real problem with computer programming.

**Example 1**: Consider the maximization of $b$ in $f(x) = ax^2 + bx + c$ where $a = c = 1$ and $x, b \in (0,1)$. Let

$$\Pi(x) = (1+x^2)^2 f\left(\frac{x^2}{1+x^2}\right) = (a+b+c)x^4 + (b+2c)x^2 + c$$

Then, the equivalent SDP problem is:

SDP2:     $Max\ y_1$

    $Subject\ to$:

$$\begin{bmatrix} y_2 & y_3 & y_4 \\ y_5 & y_6 & y_7 \\ y_8 & y_9 & y_{10} \end{bmatrix} \geqslant 0$$

$$y_2 = c = 1$$

$$y_3 + y_5 = 0$$

$$-y_1 + y_4 + y_6 + y_8 = 2c = 2$$

$$y_7 + y_9 = 0$$

$$-y_1 + y_{10} = a + c = 2$$

$$y_1 = b$$

Using an SDP software, such as SeDuMi or CVX, the optimal solution $b = 1$ is found, which can also be verified by using classical methods.

## V. Simulation Results

In this section, we present some numerical results obtained by computer simulations. Both regular and irregular parity check node degree distributions are considered. We present some points with fixed parity check degree distribution and fixed erasure probability.

## A. Some numerical results

Table 1 shows values of variable node degree distributions for some codes when parity check node degree distribution is one-tap and erasure probability is considered. These codes which obtained through SDP method are the main comparable results with the known results of pervious researches [8,18,19].

Table 1: Numerical results for rate maximization with regular parity check node

|  | $\rho(x) = x^3$ | $\rho(x) = x^4$ | $\rho(x) = x^5$ | $\rho(x) = x^6$ | $\rho(x) = x^7$ |
|---|---|---|---|---|---|
|  | **Code 1** | **Code 2** | **Code 3** | **Code 4** | **Code 5** |
| $\lambda_2$ | 0.5208 | 0.4393 | 0.4021 | 0.4387 | 0.4331 |
| $\lambda_3$ | 0.1458 | 0.2097 | 0.2137 | 0.1456 | 0.1583 |
| $\lambda_4$ |  | 0.0536 |  |  |  |
| $\lambda_5$ | 0.3333 | 0.2974 |  | 0.4158 | 0.4086 |
| $\lambda_6$ |  |  |  |  |  |
| $\lambda_7$ |  |  | 0.3902 |  |  |
| $\varepsilon$ | 0.64 | 0.56 | 0.49 | 0.38 | 0.33 |
| $R$ | 0.3346 | 0.421 | 0.4922 | 0.593 | 0.6439 |
| $C$ | 0.36 | 0.44 | 0.51 | 0.62 | 0.67 |
| $\delta$ | 0.0708 | 0.0432 | 0.0349 | 0.0435 | 0.039 |

Now, we compare Table1 with the results reported in the literature based on the following 5 criteria which proposed in [8].

1- Lower maximum degree

2- Higher rate

3- Higher threshold

4- Lower fraction of degree-two edges

5- Minimum gap to the channel capacity $\delta = 1 - R/C$

Table 2: Comparison of two codes designed previously and code 3 all with $\rho(x) = x^5$

| Reference | [18] Type-A | [18]Type-MB | Code 3 of Table 1 |
|---|---|---|---|
| $\lambda_2$ | 0.4167 | 0.4167 | 0.4021 |
| $\lambda_3$ | 0.1667 | 0.1667 | 0.2137 |
| $\lambda_4$ | 0.1000 | 0.1000 | |
| $\lambda_5$ | 0.0700 | | |
| $\lambda_6$ | 0.0532 | | |
| $\lambda_7$ | 0.0426 | | 0.3902 |
| $\lambda_8$ | 0.0353 | 0.3176 | |
| $\lambda_9$ | 0.0300 | | |
| $\lambda_{10}$ | 0.0260 | | |
| $\lambda_{11}$ | 0.0229 | | |
| $\lambda_{12}$ | 0.0204 | | |
| $\lambda_{13}$ | 0.0165 | | |
| $\varepsilon$ | 0.48 | 0.48 | 0.49 |
| $R$ | 0.4998 | 0.4926 | 0.4922 |
| $C$ | 0.52 | 0.52 | 0.51 |
| $\delta$ | 0.0389 | 0.0527 | 0.0349 |

According to the above criteria, it is seen that code 3 outperforms its competitors.

The proposed method can also be used effectively for optimizing a non-constant parity check node degree distribution. Table 3 shows such variable node and check node degree.

Table 3: Comparing a new 2-tap code with previous designed codes

|  | $\rho(x) = 0.608291x^5 + 0.391709x^6$ | $\rho(x) = 0.5x^7 + 0.5x^8$ | $\rho(x) = 0.48555x^5 + 0.51445x^6$ |
|---|---|---|---|
| Reference | [19] | [8. Example 3.63] | This Paper |
| $\lambda_2$ | 0.205031 | 0.106257 | 0.4032 |
| $\lambda_3$ | 0.455716 | 0.486659 | 0.1512 |
| $\lambda_7$ |  |  | 0.4454 |
| $\lambda_{11}$ |  | 0.010390 |  |
| $\lambda_{14}$ | 0.193248 |  |  |
| $\lambda_{15}$ | 0.146004 |  |  |
| $\lambda_{20}$ |  | 0.396694 |  |
| $\varepsilon$ | 0.5 | 0.4741 | 0.45 |
| $R$ | 0.433942 | 0.5 | 0.5267 |
| $C$ | 0.5 | 0.5259 | 0.55 |
| $\delta$ | 0.14 | 0.0493 | 0.0423 |

Based on the above 5 criteria, it is seen that our new code outperforms similar codes reported in the literature.

In order to show the efficiency of the proposed method, we discretize the interval (0,1] into a discrete set $\{x_0, x_1, \ldots, x_N\}$, for different $N$. The results are given in Fig. 1 for $\rho(x) = x^4$ and $\varepsilon = 0.56$ in which the horizontal axis shows $N$ and the vertical axis is the obtained values of $\lambda_i$s, where the obtained value for $\lambda_4$ is zero.

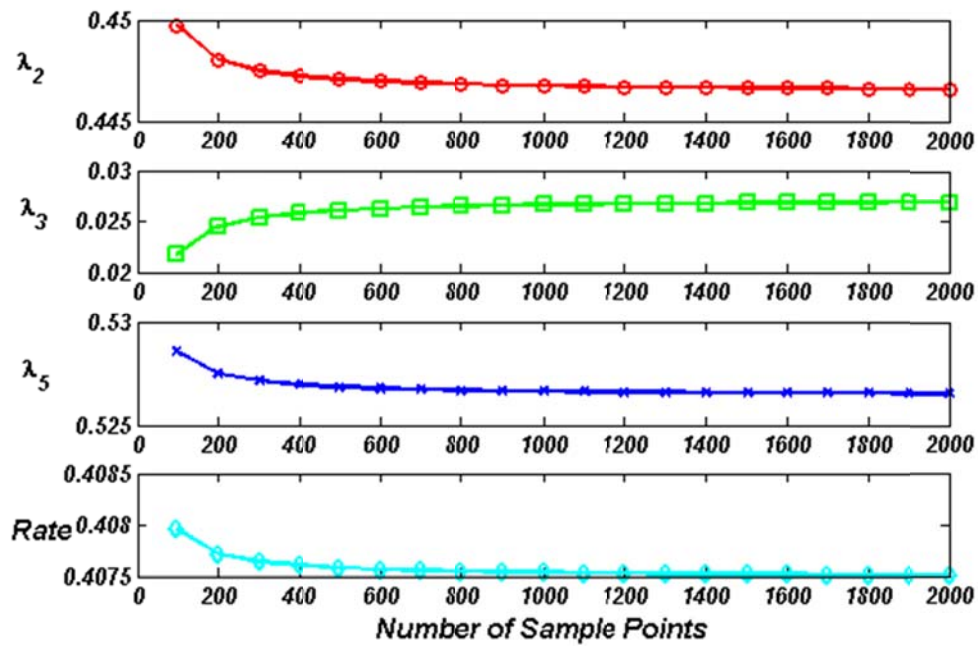

Fig.1 The found answer by using discretizing.

As it is apparent, with increasing the number of sample points, the rate and other values of $\lambda_i$'s converge to an ultimate value. For this input: $\rho(x) = x^4$ and $\varepsilon = 0.56$, Code 2 of Table 1 with $R = 0.421$ is better than the code obtained by discretizing method reported in [9].

### B. Analyzing the introduced method

Figure 2 illustrates the zero order and the first order upper bounds defined in [23] and Codes of Table 1. It is apparent that these codes achieve the upper bound.

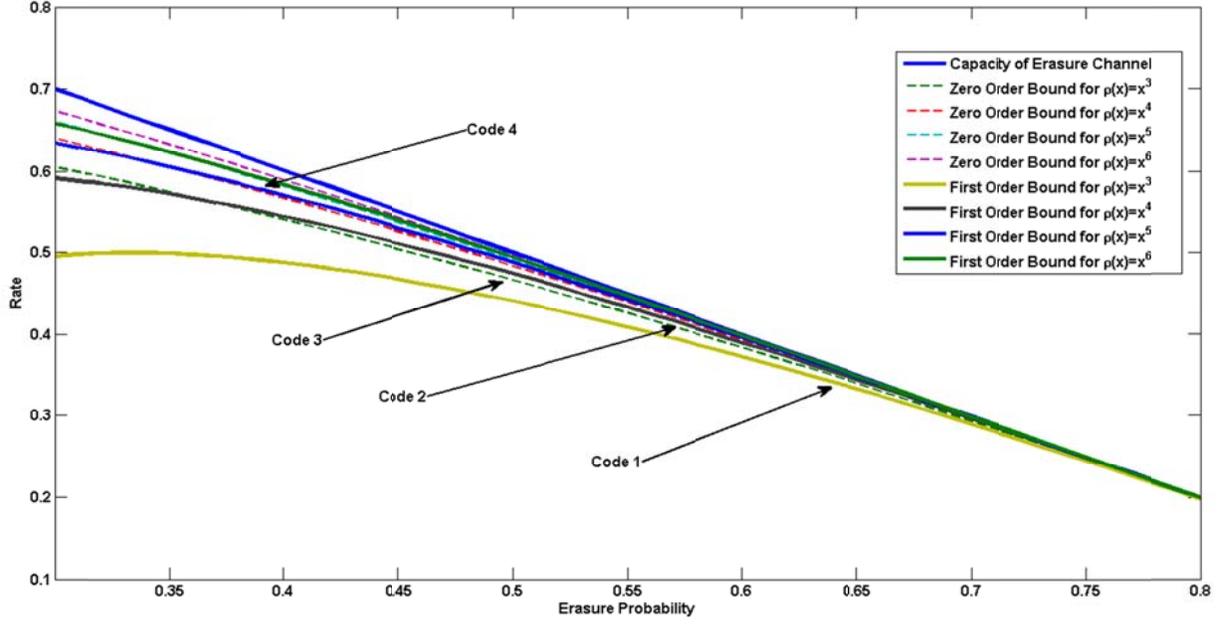

Fig 2. The upper bounds and codes optimized for $\rho(x) = x^3$, $\rho(x) = x^4$, $\rho(x) = x^5$ and $\rho(x) = x^6$. The arrows point to codes of Table 1

## VI. Conclusion

In this paper, rate optimization problem which was designed as an NLP in the literature has been modeled as a semi-definite optimization problem without any relaxation or simplification. Simulation results in both regular and irregular parity check node degree distributions were presented. These results in all cases are better than the best results reported in previous works. A wide class of irregular parity-check node degree-distribution codes may be designed by the proposed method and compared with the related bounds.

**Acknowledgement**

This work was partially supported by ITRC of Iran.